# 2D Hexagonal Boron Nitride-based Anticorrosion Coatings

Onurcan Kaya[1,2.3#], Luca Gabatel[4,5#], Sebastiano Bellani[4], Fabrizio Barberis[6], Francesco Bonaccorso[4], Ivan Cole[7] and Stephan Roche[1,8,*]

[1] Catalan Institute of Nanoscience and Nanotechnology (ICN2), CSIC and BIST, Campus UAB, Bellaterra, 08193, Barcelona, Spain
[2] School of Engineering, RMIT University, Melbourne, Victoria, 3001, Australia
[3] Department of Electronic Engineering, Universitat Autonoma de Barcelona (UAB), Campus UAB, Bellaterra, 08193 Barcelona, Spain
[4] BeDimensional S.p.A., Via Lungotorrente Secca 30R, Genova 16163, Italy
[5] Department of Mechanical, Energy, Management and Transport Engineering (DIME), Università di Genova, Genova, Italy
[6] Department of Civil, Chemical and Environmental Engineering (DICCA), Università di Genova, Genova, Italy
[7] School of Engineering, The Australian National University, Canberra, ACT ,2600
[8] ICREA Institucio Catalana de Recerca i Estudis Avancats, 08010 Barcelona, Spain
E-mail: stephan.roche@icn2.cat
[#] These 2 authors contributed equally

**Abstract**

The corrosion of metallic surfaces poses significant challenges across industries such as petroleum, energy, and biomedical sectors, leading to structural degradation, safety risks, and substantial maintenance costs. Traditional organic and metallic coatings provide some protection, but their limited durability and susceptibility to harsh environmental conditions necessitate the development of more advanced and efficient solutions. This has driven significant interest in two-dimensional (2D) materials, with graphene extensively studied for its exceptional mechanical strength and impermeability to gases and ions. However, while graphene offers short-term corrosion protection, its high electrical conductivity presents a long-term issue by promoting galvanic corrosion on metal surfaces. In contrast, hexagonal boron nitride (*h*-BN) has emerged as a promising alternative for anticorrosion coatings. *h*-BN combines exceptional chemical stability, impermeability, and electrical insulation, making it particularly suited for long-term protection in highly corrosive or high-temperature environments.

While *h*-BN holds promise as anticorrosion material, challenges such as structural defects, agglomeration of nanosheets, and poor dispersion within coatings limit its performance. This review provides a comprehensive analysis of recent advancements in addressing these challenges, including novel functionalization strategies, scalable synthesis methods, and hybrid systems that integrate *h*-BN with complementary materials. By bridging the gap between fundamental research and industrial applications, this review outlines the potential for *h*-BN to revolutionize anticorrosion technologies. These obstacles necessitate advanced strategies such as surface functionalization to improve compatibility with polymer matrices and dispersion optimization to minimize agglomeration. Recent advancements highlight the incorporation of *h*-BN into composite materials, which have shown significant advances in durability, adhesion, and overall performance.

Future directions for *h*-BN research emphasize scalable fabrication techniques to produce large-area, defect-free coatings suitable for industrial deployment. Furthermore, hybrid systems that integrate *h*-BN with complementary materials are proposed to enhance corrosion resistance and address specific environmental and operational demands. These approaches hold the potential to establish *h*-BN as a transformative material for next-generation anticorrosion technologies.

## 1. Introduction

The effect of corrosion of metallic parts and surfaces is a significant concern across various industries, including petroleum, energy, and biomedical applications [1–3]. Metal components used in these industries are frequently exposed to harsh environmental conditions such as fluctuating temperatures, high humidity and chemical pollutants. These factors can accelerate the corrosion process, leading to material degradation through chemical and electrochemical reactions [3–6]. This not only compromises the structural integrity and operational safety of equipment but also incurs substantial economic costs due to increased maintenance, repairs and replacements. Hou et al. [2] estimated that the cost of corrosion is ~310 billion USD in China alone, while Koch [1] estimated the global cost of corrosion at ~2.5 trillion USD, accounting for 3.4% of the global Gross Domestic Product (GDP).

Traditional organic and metallic coatings offer limited durability under harsh conditions, driving the search for advanced materials. Recently, two-dimensional (2D) materials [7] have garnered considerable attention for their potential use in anticorrosion coatings. Among these, graphene has been extensively studied due to its excellent mechanical properties and well-known impermeability to gases and ions [8,9]. Early studies by Chen *et al*. [10] and Prasai *et al*. [11] demonstrated that pristine graphene layers could effectively reduce the corrosion rates of metal surfaces. However, the high electrical conductivity of graphene often exacerbates galvanic corrosion, particularly in presence of defects [12–14]. This realization has shifted focus toward insulating 2D materials such as hexagonal boron nitride (*h*-BN), which offers a compelling alternative due to its chemical stability and electrical insulation properties. Schriver *et al*. [15] and other research groups [16,17] revealed that graphene is not ideal for long term corrosion protection due to its high conductivity and electrochemical nobility, which may lead to galvanic and pitting corrosion when coupled with most common metallic materials [16,17]. Galvanic corrosion occurs when two metallic materials with different corrosion potentials come into contact in the presence of a conductive medium, such as humid air or an aqueous electrolyte. This results in the preferential oxidation of the anodic material while protecting the cathodic material. Even when protective coatings are applied, defects within the coating can allow the electrolyte to penetrate, forming an electrochemical pathway that accelerates the corrosion process. For instance, conductive coatings such as graphene, despite their excellent barrier properties, can promote galvanic corrosion by acting as a cathode in this circuit. The use of insulating materials, such as *h*-BN, can prevent the formation of such electrochemical pathways and improve long-term corrosion resistance [15,18,19].

*h*-BN, with its hexagonal lattice structure, offers excellent chemical stability and impermeability to ions and moisture, similar to graphene [20,21]. However, unlike graphene, *h*-BN is electrically insulating and exhibits very low electrochemical activity [20,22]. Its insulating nature prevents the formation of galvanic cells, a primary cause of accelerated corrosion in long-term exposure. Furthermore, its hydrophobic properties help repelling water and other corrosive agents, reducing the corrosion of metal surfaces [20,22–25]. Despite its advantages, *h*-BN faces several challenges before it can be effectively used as a coating material. Defects, wrinkles and other structural imperfections that arise during the fabrication process of the 2D nanosheets can compromise its protective properties, creating pathways for corrosive agents and leading to localized areas where corrosion may initiate and propagate [26–28]. Furthermore, producing high-quality, defect-free *h*-BN nanosheets at large scale is challenging and expensive, limiting its economic viability for mass production [24,29]. The hydrophobic nature of *h*-BN can also cause the nanosheets agglomeration in composite coatings prepared from polar solvent-based paints, reducing their barrier effectiveness. Moreover, *h*-BN may not be inherently compatible with certain polymers and other materials, leading to poor adhesion, delamination or peeling of the coating [21,30]. Researchers are exploring various approaches to address these issues, including surface functionalization, advanced dispersion techniques, and hybrid coating strategies to enhance the performance and viability of *h*-BN-based anticorrosion coatings.

This review explores the potential of *h*-BN-based coatings for anticorrosion applications, focusing on mechanisms by which *h*-BN and other 2D materials protect metal surfaces, such as physical barriers, self-healing, and hydrophobicity (Section 2). Section 3 addresses the strategies developed by researchers to overcome challenges, including issues related to dispersion, structural defects, and compatibility with polymers. Section 4 highlights the latest technological applications and fabrication techniques for *h*-BN-based coatings with an emphasis on industrial exploitation. Finally, Section 5 outlines future research directions and potential advancements to enhance the effectiveness and commercial viability of *h*-BN-based anticorrosion coatings.

## 2. Protection and failure mechanisms of 2D materials in anticorrosion coatings

An ideal anticorrosion coating material should be highly impermeable to ions and corrosive agents while maintaining long-term stability without cracking, even under harsh conditions [30,31]. The exceptional impermeability and chemical stability of 2D materials make them excellent candidates for such applications. In addition, they enhance the anticorrosion performance of polymer-based composite coatings, which are extensively used in real-world applications [9,32]. By blocking corrosive agents or extending their diffusion pathways, 2D materials can delay the onset of corrosion. Furthermore, they fill pores and cracks in the polymer matrix, improving coating performance [7]. Beyond impermeability, 2D materials can enhance hydrophobicity and adjust the electrical conductivity of the anticorrosion coatings, playing a key role depending on the final use, *e.g.*, in a primer or topcoat. This section examines these protective mechanisms, including physical barrier effect, self-healing properties, hydrophobicity, and the role of electrical conductivity in corrosion prevention.

### 2.1. *Physical barrier effect*

A pristine, defect-free 2D material layer is theoretically impermeable to corrosive agents, making it an effective protective barrier. 2D materials can either be directly applied to the metal surface to act as a shield against corrosive media or incorporated into polymer coatings to create a "maze effect" by extending and torturing the diffusion pathways of the corrosive substances [33–35]. The impermeability of 2D materials stems from their atomic structure. For example, graphene has a lattice diameter of 2.46 Å, but due to the length of the C-C bonds (1.42 Å) and the van der Waals radius of C atoms, the pore diameter of the graphene lattice is reduced to 0.64 Å, effectively blocking gas molecules or corrosive ions [8,31,36]. Additionally, the delocalized and dense electron clouds of $\pi$- $\pi$ bonds block the gap within the C rings, creating a physical separation between the metal surface and corrosive agents by repelling reactive atoms and molecules [12,37–39]. This high energy barrier prevents the diffusion of corrosive agents to the underlying metal surface [12,37–39]. Early studies demonstrated exceptional impermeability of chemical vapour deposition (CVD) grown graphene and significantly reduced corrosion rates when it is applied as a coating for metal surfaces such as carbon steel, NiTi alloy and copper surfaces [10,11,40,41]. However, practical implementation of graphene presents major challenges. Producing continuous, defect-free graphene sheets or any 2D material on a large scale remains a major challenge for industrial development [42,43].

Wrinkles, oxidation and other defects introduced during the fabrication or transfer of graphene onto the metallic surface can create pathways for corrosive agents to reach the substrate to be protected [9,36,44,45]. Additionally, the high carrier mobility of graphene, which is also electrochemically noble, and other conductive 2D materials can promote galvanic corrosion in long-term applications [15,46]. These limitations have led to increasing interest in electrically insulating alternatives such as *h*-BN.

Unlike graphene, *h*-BN sheets provides a robust barrier against corrosive species while being electrically insulating, eliminating galvanic corrosion risks. Shen *et al*. [47] demonstrated that both graphene and *h*-BN sheets impose similar energy barriers against oxygen penetration, reinforcing their potential as protective layers. However, *h*-BN exhibits superior chemical stability in oxidative environments. [48] Liu *et al*. [48] have shown that *h*-BN sheets can stop oxidation even at temperatures as high as 1100 ºC thanks to the high energy barrier to oxygen absorption, thus providing an exceptional protection to nickel surfaces. A more detailed atomic-scale understanding of the oxidation resistance of *h*-BN is provided by Zhao et al. [49], which reveals that oxygen adsorption on *h*-BN is highly site-selective. Oxygen atoms preferentially migrate toward nitrogen sites due to their higher electronegativity, but once in proximity to boron atoms, they experience strong binding interactions that restrict further diffusion. This results in localized oxygen trapping rather than uncontrolled oxidation, which prevents the formation of continuous oxide layers that could degrade the structural integrity of the coating. Additionally, the calculations show that oxygen-induced structural distortions in *h*-BN are significantly lower than those in graphene, further reinforcing its superior resistance to oxidative environments [49]. Beyond oxidation, *h*-BN also protects against other corrosive species. [23] Chilkoor *et al*. demonstrated the effectiveness of *h*-BN against microbial [23] and sulfur-induced degradation [50], highlighting its broader stability in aggressive chemical environments. Nevertheless, like graphene, *h*-BN coatings are susceptible to structural defects such as wrinkles, misaligned grain boundaries, and adhesion challenges, which limit their scalability for large-area corrosion protection [15,19,47]. These limitations have led to an alternative strategy, incorporating 2D materials into polymer matrices to reinforce coatings and extend the durability of corrosion-resistant barriers.

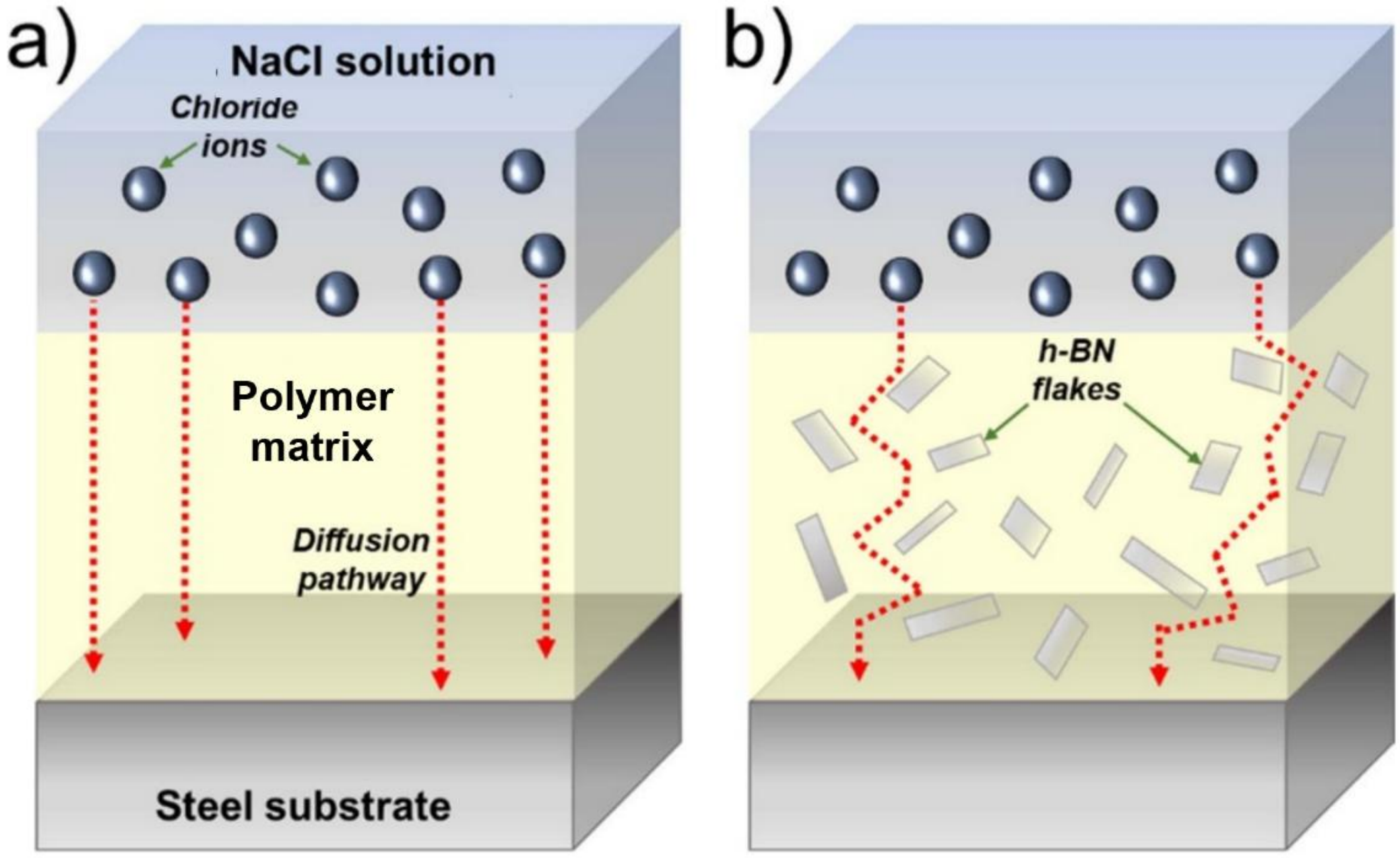

***Figure 1***: *The diffusion pathways for corrosive ions (a) and the extended pathways created by impermeable h-BN flakes (b) within the polymer matrix [29].*

Polymeric coatings are widely used for corrosion protection due to their chemical versatility, flexibility, and ease of application. However, their long-term stability is inherently limited, as they inevitably develop structural defects over time. During the curing process, solvent evaporation leads to micropores and cracks, which act as direct penetration channels for moisture, oxygen, and chloride ions, accelerating metal corrosion [9,32,51,52]. These defects can also emerge from differential thermal expansion between the coating and the metal substrate, causing internal stress accumulation that further degrades adhesion and structural integrity. Additionally, mechanical wear, UV exposure, and environmental stressors progressively weaken the polymer matrix, leading to the formation of permeation pathways that allow corrosive species to reach the underlying metal [32,51,52].

The incorporation of 2D materials, such as graphene and *h*-BN, into polymer coatings can alter the transport mechanisms of corrosive species, enhancing the protective performance of these coatings. Rather than allowing direct penetration, these materials disrupt the linear diffusion path, forcing corrosive agents to navigate a highly tortuous and extended route, significantly delaying their breakthrough time, as illustrated in **Fig. 1** [53]. Du et al. [54] experimentally demonstrated that that ultrathin ionic liquid-functionalized *h*-BN nanosheets, incorporated into epoxy (EP) coatings, significantly increased the diffusion resistance by creating a dense, well-dispersed barrier structure. Their electrochemical measurements showed that coatings with 0.5 wt% *h*-BN maintained a high resistance ($5.6 \times 10^9$ Ω $cm^2$) even after 4 weeks of immersion in 3.5 wt% NaCl solution, which is four orders of magnitude higher than that of pure epoxy coatings. Mujib et al. [35] further demonstrated that the incorporation of *h*-BN can reduce ion permeability by up to two orders of magnitude, reinforcing the ability of *h*-BN to restrict corrosive species transport in various coating formulations.

The effectiveness of 2D materials as physical barriers in polymer coatings is strongly influenced by their adhesion to the polymer matrix, as well as their dispersion and orientation. Strong adhesion ensures that the nanosheets remain well-integrated within the coating, preventing delamination and the formation of interfacial defects that could serve as entry points for corrosive species [31,55]. Adhesion between 2D nanosheets and polymers is governed by intermolecular interactions such as van der Waals forces, hydrogen bonding, dipole-dipole interactions, and mechanical interlocking at the interface [31,55]. However, many 2D materials, including pristine *h*-BN and graphene, have weak interfacial adhesion due to their low surface energy and chemical inertness, making them prone to poor dispersion and weak bonding with polymer matrices [31,55]. To overcome this, functionalization strategies, such as the introduction of oxygen-containing groups, silane modifications, or ionic liquid-assisted processing, can improve interfacial interactions, enhancing adhesion and overall barrier performance [56–58]. *h*-BN, despite its weak surface interactions, can still enhance barrier properties when well-integrated, as shown by Du et al. [54], where ionic liquid-assisted *h*-BN coatings exhibited superior interfacial stability and significantly higher corrosion resistance compared to pure epoxy. Similarly, Sun et al. [57] found that silicon-modified reduced graphene oxide (RGO), enhances interfacial adhesion, creating a more compact and defect-free barrier that resists delamination [57]. Detailed discussion of these functionalization approaches is presented in Section 3.2.

Dispersion of 2D materials in composite coatings is another critical factor, as poor dispersion leads to agglomeration, which weakens the physical barrier effect by forming localized defects where corrosive species can directly penetrate [59–62]. This occurs since most pristine 2D materials, including *h*-BN, tend to re-stack due to strong van der Waals forces and poor wettability with polymer matrices, making uniform dispersion challenging [63]. A homogeneously dispersed network of nanosheets forces corrosive agents to take a more tortuous diffusion path, increasing the overall coating resistance to corrosion agents [64]. In addition to dispersion, orientation further improves barrier efficiency. In the case of graphene, randomly distributed flakes may contact the metal and form a galvanic cell, which does not occur in a highly ordered structure. Moreover, the highly oriented structure can maximize the physical barrier effect. Ding *et al*. [65] demonstrated that aligning carbon dot-functionalized $Ti_3C_2T_x$ nanosheets in epoxy coatings through an air-flow-induced self-assembly method can prevent galvanic corrosion while maintaining a defect-free barrier. Similarly, Fan *et al*. [66] used electric fields to orient $Ti_3C_2T_x$ nanosheets in epoxy, enhancing the anticorrosion performance and extending the coating lifetime. Wang *et al*. [67] demonstrated that the orientation and arrangement of the graphene layers within the polymers can affect the anticorrosion performance.

### 2.2. *Self-healing*

The self-healing capability of the coatings is crucial for long-term application, as coatings without this ability may fail to safeguard the underlying metal surfaces after sustaining damage. Polymers can exhibit self-healing through intrinsic and extrinsic mechanisms [68,69]. Intrinsic healing relies on the reversible repair of the polymer matrix via dynamic covalent or non-covalent bonds [68,69]. Reversible covalent bonds require significant energy and external stimuli, such as heating or light, to form and break [68,69]. In contrast, non-covalent bonds, such as hydrogen bonds, π-π interactions, and host-guest interactions, require less energy and can occur even at room temperature. Extrinsic healing involves incorporating repair materials in the damaged coating. These materials can either inhibit corrosion in the damaged area or fill and repair the cracks in the coating. Besides providing a physical barrier, 2D materials can enhance both intrinsic and extrinsic healing mechanisms within the polymer matrix [68,69].

One effective method for intrinsic healing using covalent bonds involves heating the coating to facilitate thermally reversible Diels-Alder reactions [70–72]. This process is particularly relevant in coatings containing furan-maleimide functional groups, which undergo dynamic covalent bond rearrangement under heat. Cao *et al*. demonstrated this effect in an *h*-BN/epoxy composite coating, where maleimide-functionalized *h*-BN nanosheets participated in the reaction. Upon heating to 120°C, the composite coating successfully healed microcracks, restoring its protective function [71,72]. In these applications, efficient thermal transfer to the polymer matrix is crucial. In these studies, due to its high thermal conductivity and stability, *h*-BN sheets act as a thermal conductor, ensuring even heat distribution throughout the polymer network [71,72]. Graphene can also be used in a similar fashion. Beyond that, graphene exhibits strong microwave and IR adsorption due to its high electrical conductivity. This allows it to be heated by microwave or light and efficiently transfer heat to the polymer matrix, facilitating the reversible formation and breaking of chemical bonds [68,70]. While D–A-based self-healing relies on covalent bond dynamics, self-healing can also be achieved through non-covalent interactions, such as metal coordination and π–π stacking. Li *et al*. [73] demonstrated this approach by developing a superhydrophobic anticorrosion coating with self-healing properties using polydopamine (pDA)-functionalized $Cu^{2+}$-decorated GO nanosheets, octadecylamine (ODA), and polydimethylsiloxane (PDMS). Unlike D–A reactions, which require external heating, this system repairs itself spontaneously at room temperature through π–π interactions between GO and pDA and chelation between catechol groups and $Cu^{2+}$ ions. These non-covalent interactions enable defect healing without additional energy input, while the hydrophobicity of ODA and PDMS further reinforces the barrier against moisture and chloride ions. After scratching and immersing the coatings in salt water for 30 days, the system exhibited remarkable anticorrosion performance, demonstrating its long-term stability in aggressive environments [73].

Functionalized 2D materials play a critical role in extrinsic self-healing mechanisms, acting as nanocontainers that store and release active agents upon damage. Graphene oxide (GO), for example, has been widely studied for its ability to adsorb ions and functionalized molecules, allowing it to serve as a smart reservoir for corrosion inhibitors [53,74]. Similarly, mesoporous silica nanoparticles (MSNs) have been incorporated into polymer coatings to control the release of healing agents, enabling responsive and long-lasting corrosion protection [75,76] The combination of GO and MSNs has gained particular attention due to its ability to provide both active and passive protection, reinforcing coating integrity while delivering targeted repair agents upon environmental triggers [53,74,77]. The effectiveness of these nanocontainer-based systems is largely dependent on their stimuli-responsive release behaviour. Ma *et al.* [53] demonstrated that benzotriazole (BTA)-loaded MSNs functionalized with GO exhibited pH-sensitive release, activating when the coating was damaged. As corrosion initiates, the resulting chemical reactions alter the local pH, triggering the controlled release of BTA, which suppresses further metal oxidation and repairs the protective layer. This self-regulating mechanism ensures that inhibitors are only deployed when necessary, extending the coating's lifespan and efficiency. The mechanism of BTA release in response to pH changes is illustrated in **Fig. 2** [53]. Building upon this approach, Wang *et al.* [75] developed a dual-responsive healing system by functionalizing amine-modified MSNs with GO and loading them with tannic acid (TA), a corrosion inhibitor. When incorporated into an epoxy matrix, this system was tested under alternating hydrostatic pressure conditions, a significant challenge in deep-sea applications. Their results demonstrated that the self-healing coating maintained exceptional anticorrosion performance, even in harsh underwater environments, highlighting the potential of these systems for marine and industrial applications [75].

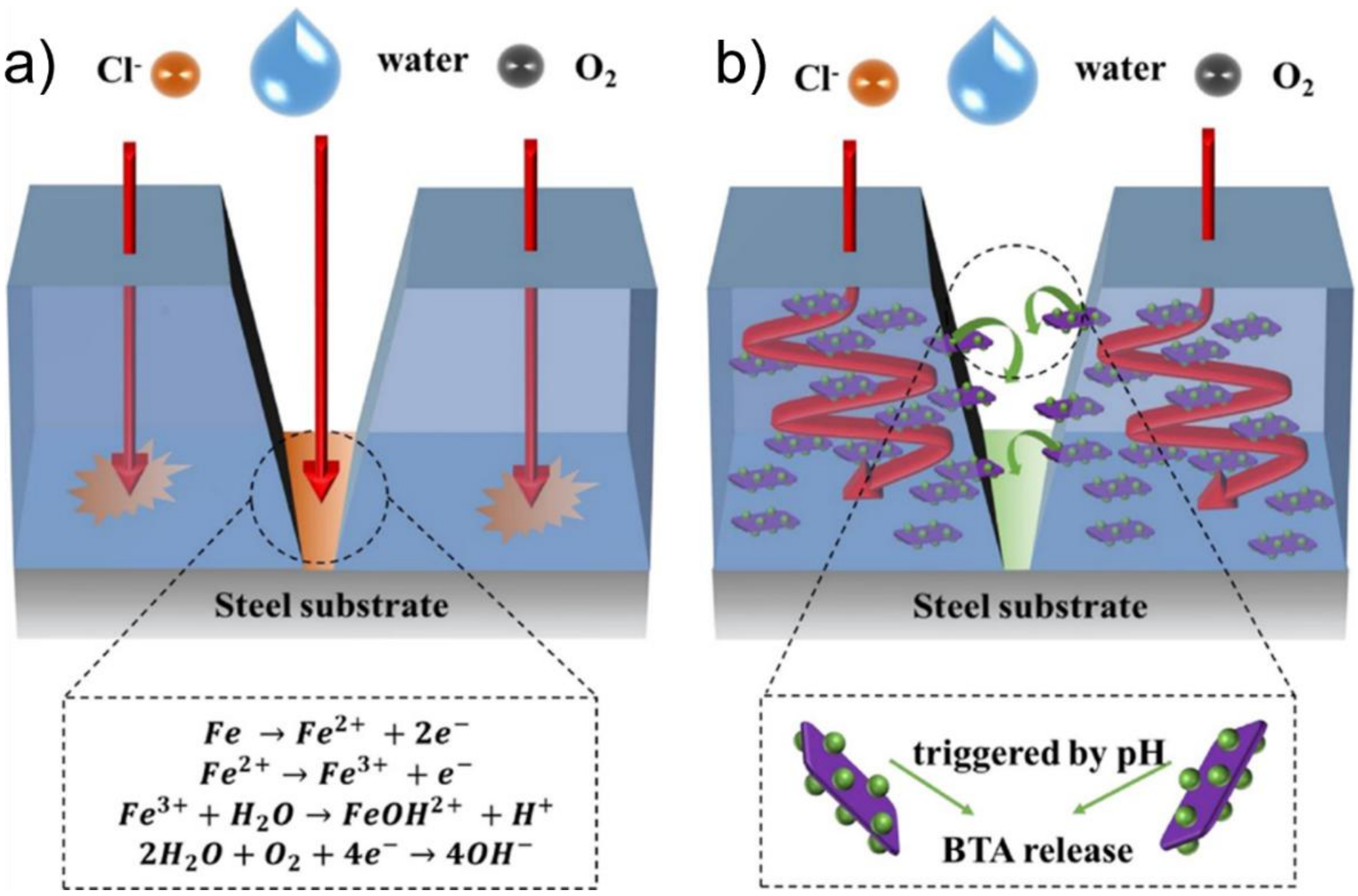

***Figure 2**. Schematic presentation of the anticorrosion performance of a) blank waterborne epoxy (WEP) and b) fGO/WEP [53].*

## 2.3. *Hydrophobicity*

Hydrophobicity is an important property for anticorrosion applications because it minimizes interaction with corrosive species, such as water and ions, thereby reducing the corrosion rate [78–81]. A surface is considered hydrophobic when its water contact angle exceeds 90º and superhydrophobic when it surpasses 150º, causing water droplets to roll off, preventing moisture accumulation [81]. This behaviour is influenced by surface chemistry, which determines surface free energy, and microstructural properties, such as roughness and nanoscale morphology, which influence wetting behaviour. Two-dimensional (2D) materials have been extensively explored for their ability to enhance hydrophobicity in coatings due to their low surface energy, tunable wettability, and ability to modify surface topography. Their incorporation improves barrier properties by reducing water permeability and influencing how water interacts with the coating. However, their effectiveness depends on a combination of intrinsic hydrophobicity, structural arrangement, and dispersion stability [55,78–80].

Certain 2D materials, such as hexagonal boron nitride (*h*-BN) and graphene, exhibit inherent hydrophobicity due to their non-polar surfaces, resulting in high water contact angles and minimal adhesion to water molecules. Their hydrophobic nature reduces water absorption, thereby preventing direct contact between the coating and corrosive species [29,78,82]. Tang *et al*. used *h*-BN films directly grown on SS304 steel to create a highly hydrophobic surface, achieving excellent corrosion resistance over 10 weeks [82]. Molina-Garcia *et al*. [29] reported that adding 10 wt% wet-jet milled *h*-BN into a polyisobutylene (PIB) matrix increased from approximately 88º to over 100º with an order of magnitude reduction in corrosion rate [29]. While *h*-BN and graphene are intrinsically hydrophobic, their properties can be further improved through surface functionalization and chemical modifications. Wang *et al*. [63] used polystyrene (PS) microspheres to decorate *h*-BN nanosheets, successfully preventing their coalescence. Cui *et al*. [83] modified *h*-BN nanosheets using π-π interactions of amine-capped aniline trimer to improve the stability of the *h*-BN nanosheets in the polymer matrix. Sun *et al*. [78] used urushiol (Ur) to prevent agglomeration of *h*-BN nanosheets in epoxy coatings. Ur-functionalized *h*-BN nanosheets exhibited excellent hydrophobicity, which improved long-term corrosion prevention performance. Wu *et al*. [84] functionalized *h*-BN nanosheets with amphiphilic GO films via π-π interactions, improving the dispersibility of *h*-BN films, while preserving the required hydrophobicity in the coatings.

Functionalization techniques can convert intrinsically hydrophilic surfaces, such as GO and MXenes, into hydrophobic coatings. Their oxygen-rich functional groups promote strong hydrogen bonding with water, making them naturally hydrophilic. However, chemical modifications, such as fluorination, silanization, or polymer grafting, lower surface energy and shift these materials toward a hydrophobic or superhydrophobic behaviour [33,34,85]. Zhang *et al*. [86] enhanced the wettability of $Ti_3C_2T_x$ nanosheets by functionalizing them with [3-(2-Aminoethyl)aminopropyl] trimethoxysilane (AEAPTES), a silane coupling agent. This modification resulted in a significant increase in the water contact angle from 6º–8.5º to 136º–142º, approaching superhydrophobic levels. Similarly, Dong *et al*. [79] perfluorodecylsilane (PFDS) functionalization of GO nanosheets increased the contact angle from 9º to 146º, improving water repellency and corrosion resistance. However, excessive functionalization or filler loading beyond 1 wt% led to agglomeration, creating defects that compromised corrosion protection [79].

Hydrophobicity enhancement is not solely dependent on chemical modifications; the interaction of 2D materials with the coating matrix, their dispersion, and their structural arrangement strongly influence wetting behaviour and water repellency. Surface roughness plays a critical role in repelling moisture by trapping air pockets at the solid–liquid interface, thereby minimizing direct contact between the coating and water droplets. However, the effect of roughness is highly dependent on the orientation of 2D nanosheets within the coating. Horizontally aligned nanosheets create a continuous, low-energy barrier that hinders water penetration, but alone, they may not achieve superhydrophobicity without additional surface modifications. In contrast, vertically or partially vertically aligned nanosheets introduce hierarchical roughness, increasing air entrapment and enhancing water repellency. This structural arrangement further disrupts the formation of continuous water films, preventing ion diffusion and improving corrosion resistance. Studies have demonstrated that vertically oriented *h*-BN structures can exhibit superhydrophobic contact angles exceeding 150º, offering superior moisture protection. [82]Pakdel *et al*. [87,88] demonstrated that chemically vapor deposited (CVD) vertically aligned *h*-BN nanosheets improved surface roughness, leading to superhydrophobicity with exceptionally high water contact angles and enhanced anticorrosion performance.

### 2.4. *Electrical properties of 2D material*

As detailed earlier, galvanic corrosion remains a primary challenge in the design of protective coatings, particularly for conductive materials such as graphene. In the case of graphene-based protective coatings, the high charge transport ability of graphene can direct these small currents to the other parts of the structure, potentially accelerating the corrosion rate of the metal underneath [15,31,45]. This makes graphene unsuitable for long-term applications [15,31,45]. Similarly, MXenes and other emerging conductive 2D materials face the same issue [89]. This mechanism highlights the long-term challenges associated with using pristine graphene coatings for corrosion protection.

When graphene is in direct contact with a metal substrate, it forms a galvanic couple where the metal serves as the anode and graphene as the cathode, leading to preferential oxidation of the metal. This issue is exacerbated by defects, grain boundaries, or pinholes in the graphene layer, which allow electrolyte penetration and accelerate localized corrosion. Studies by Schriver et al. [15] and others [18,31,45] have shown that despite its excellent barrier properties, graphene coatings can indeed increase corrosion rates over extended exposure periods. MXenes face similar challenges [34,90], as their high metallic conductivity (~$10^3$–$10^5$ S/m) promotes localized electrochemical activity when exposed to electrolytes.

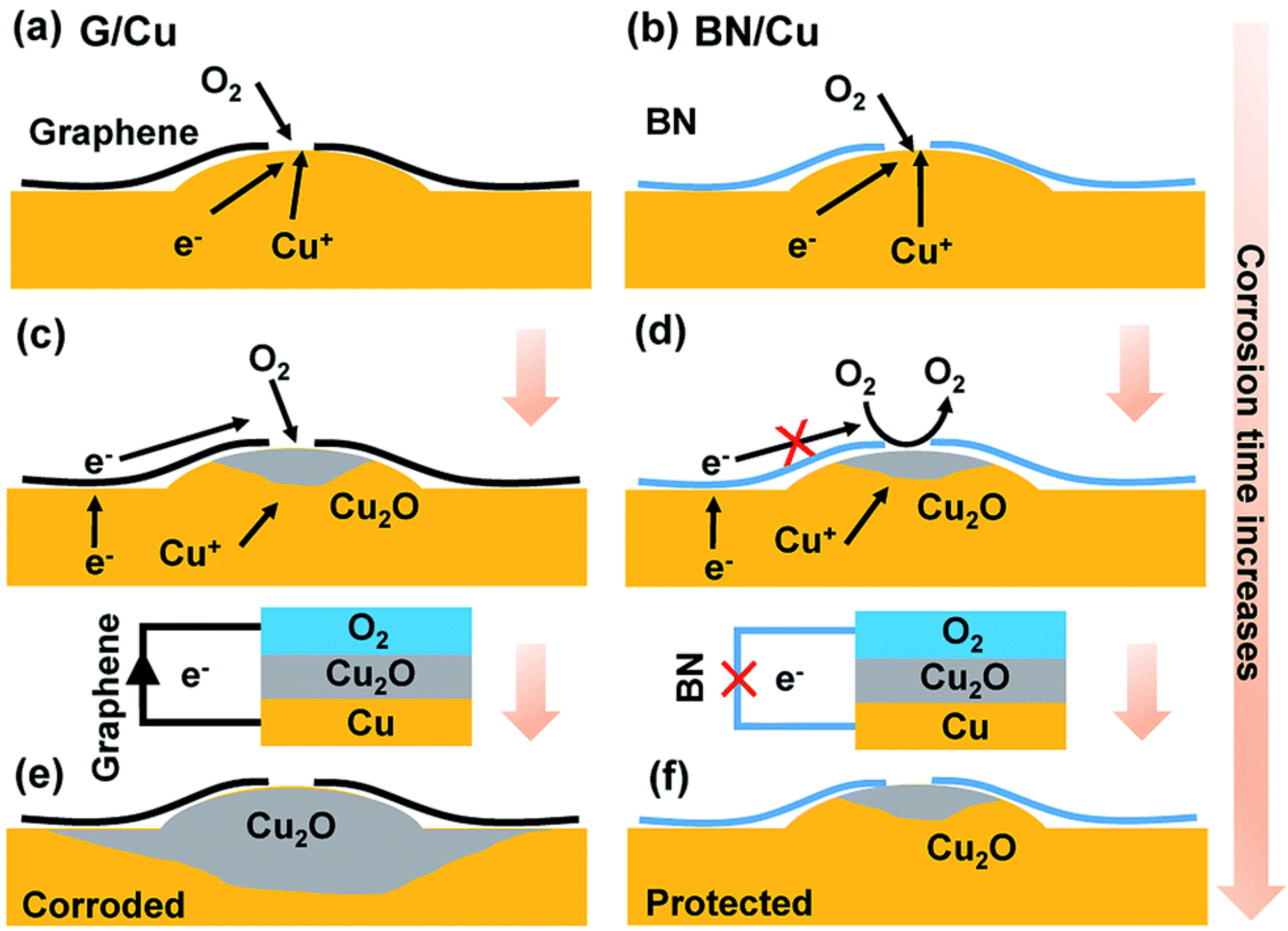

***Figure 3**: Schematic diagram illustrating the corrosion mechanisms for Cu surfaces coated with graphene and h-BN. Defected sites on (a) graphene and (b) h-BN layers, permitting oxygen to interact with the Cu surface. (c) Role of the graphene, which acts as a cathode in the electrochemical circuit due to its high conductivity and electrochemical nobility, thereby accelerating the corrosion. (d) The role of h-BN, which opens the electrochemical circuit of the h-BN-coated Cu system due to the insulating behavior of h-BN, resulting in more localized and less severe corrosion. The equivalent circuits of the (e) graphene- and (f) h-BN-coated Cu are also given [47].*

One approach to mitigate galvanic corrosion issues associated with graphene is to incorporate it into polymer matrices [9,60,67]. In such systems, the insulating nature of the polymer prevents graphene flakes from forming conductive pathways, reducing the risk of corrosion while maintaining the impermeability of graphene. However, the effectiveness of this strategy is highly dependent on filler loading [30,32]. At low concentrations (<1 wt%) [91], the polymer matrix preserves its insulating properties while extending the diffusion path for corrosive agents. As the graphene loading increases, however, the risk of electrical percolation rises, potentially leading to conductive pathways that facilitate corrosion rather than prevent it [30,31,57,91]. A similar trade-off exists for MXene-polymer composites, where careful optimization of the filler concentration is required to balance the barrier effect against the risk of conductivity-driven degradation [34,92].

The risk of galvanic corrosion has heightened interest in insulating 2D materials, especially *h*-BN [22,30]. With its large electronic band gap (~5.9 eV [93]), *h*-BN does not conduct electrons, preventing galvanic coupling with the metal substrate even in the presence of defects or pinholes. Unlike graphene, *h*-BN disrupts the electrochemical circuit necessary for corrosion currents, as depicted in **Fig. 3** [47]. Studies have demonstrated that monolayer *h*-BN coatings can significantly reduce corrosion current density by orders of magnitude across various metal substrates, even under harsh environmental conditions [23,47,50]. In addition to direct coatings, *h*-BN is widely used in polymeric matrices to enhance corrosion resistance. Unlike graphene, which requires precise filler control to prevent electrical percolation, *h*-BN allows for higher loading concentrations without introducing unwanted conductivity [30,55]. However, similarly to graphene, its hydrophobic nature may limit dispersion in certain polymer matrices, requiring functionalization techniques such as hydroxylation, amination, or surface grafting to improve compatibility [22,32,55,82].

## 3. Hexagonal boron nitride in anticorrosion coatings

This section explores the application of *h*-BN nanosheets and films in anticorrosion coatings, highlighting their material properties and key optimization strategies. It will examine the material properties of *h*-BN that contribute to its effectiveness as an anticorrosion coating and discuss key optimization strategies to address current challenges. These challenges include mitigating accelerated corrosion resulting from defects or boundaries within *h*-BN nanosheets and films, ensuring the homogeneous dispersion of them within composite coating matrices. A comprehensive understanding of the strengths and weaknesses of *h*-BN will guide its advancement as a promising 2D material for protecting metallic surfaces from corrosion.

### 3.1. ***Hexagonal boron nitride nanosheets***

As discussed in Section 2.4, the high conductivity and tendency to promote galvanic corrosion of graphene and conductive 2D materials [15] have spurred interest in electrically insulating *h*-BN [30,94]. Beyond its electrically insulating behavior, *h*-BN offers superior thermal stability, remaining intact at temperatures up to 850ºC compared to oxidation temperature of graphene (450ºC) in air [95]. This thermal resilience, coupled with its impermeability, makes it a promising candidate for protecting metal surfaces from oxidation [20,94], microbial attacks [23], and corrosive agents such as water and ions [50,96,97].

Although the results reported are encouraging, they are still based on short-term studies, with maximum durations of only a few hours. In real-world applications, longer-term protection is essential. One of the primary challenges is the presence of structural defects, such as cracks and grain boundaries, which can facilitate the penetration of corrosive agents and compromise barrier properties [98]. To improve the protective performance of *h*-BN, researchers have explored two main strategies, direct growth of defect-free *h*-BN films on metal surfaces and surface functionalization to modify interfacial interactions and passivate defects [22,54]. Khan *et al*. [99] used both experimental techniques and density functional theory (DFT) simulations to investigate the role of defects, discovering that the dissociation of moisture into $O^{-2}$ and $2H^{+}$ ions is the most energetically favourable pathway. The presence of defects, particularly around B atoms at the edges, may lead to an accumulation of $O^{-2}$ ions and potentially accelerate the corrosion rate. Experimental studies have also identified point defects, vacancies and other types of impurities in *h*-BN layers, as characteristics that impact negatively the coating performance [26–28,30,99]. Therefore, the quality of *h*-BN layers is crucial for their successful use in anticorrosion applications. A major source of structural defects arises from the transfer process used to apply *h*-BN films onto metal surfaces. Growing *h*-BN on a separate substrate before transferring it to a metal can introduce wrinkles, cracks, and contamination, which lower the effectiveness of the coating [42]. To mitigate this, researchers have focused on direct growth of *h*-BN films on metals using CVD, which minimizes defects and improves coating performance [100–105]. Tang *et al.* [82] demonstrated that CVD- grown large-area *h*-BN nanofilms, 200 nm thick and directly deposited on stainless steel, exhibited excellent anticorrosion performance over 10 weeks by acting as a physical barrier and providing hydrophobicity. Similarly, Wang *et al.* [106] reported outstanding antioxidation and anticorrosion performance for *h*-BN films grown by CVD under atmospheric conditions, noting that regions of higher quality have shown improved hydrophobicity, with larger water contact angles.

### 3.2. *Hexagonal boron nitride-based composite coatings*

The use of *h*-BN monolayers for anticorrosion applications faces several obstacles, mostly related to the scalability, cost, and quality, as discussed in Section 3.1. These issues limit their practical applicability. In contrast, polymers are often preferred in industrial applications due to their ease of processing, good barrier properties, low cost and ability to adhere well to metallic surfaces [32,107,108]. However, during the production and service life of polymer coatings, cracks and pores can be formed [32]. These defects create pathways for corrosive agents to reach the metal surface and trigger corrosion [29,109,110]. Incorporating *h*-BN or other 2D materials as fillers in polymer matrices offers a promising solution. In this context, the production of *h*-BN nanosheets by liquid-phase exfoliation (LPE) of their bulk (*i.e.*, layered) counterparts represents the ideal methodology to produce high-quality materials [7,29,42,43,111,112]. While *h*-BN monolayers act as a physical barrier or heal the cracks and pores within the polymer matrix, the latter *per se* can lower the cost of the coating application and reduce the need for large-area high-quality *h*-BN films.

In polymer coating, the impermeability of *h*-BN nanosheets prevents corrosive agents from passing through, thereby increasing the path these agents must traverse to reach the metallic surface, as illustrated in **Fig. 1**. Molina-Garcia *et al*. [29] produced high-quality few-layer *h*-BN using wet-jet milling exfoliation and screened different content of *h*-BN nanosheets in PIB-based anticorrosion coatings for marine applications. The authors found that a 5 wt% *h*-BN content in PIB coatings provided the optimal anticorrosion performance due to enhanced hydrophobicity and superior tortuosity of the paths of corrosive agents [29]. Notably, the corrosion rate was reduced to 7.4 nm/year, two orders of magnitude lower than that of pristine PIB-coated steel in a 3.5 wt% NaCl aqueous solution [29]. Similarly, Husain *et al.* used exfoliated *h*-BN flakes as a filler for polyvinyl alcohol-based composite applied to stainless steel. Their immersion tests resulted in a corrosion rate of 1.3 μm/year [113]. While *h*-BN/polymer composite coatings have consistently demonstrated enhanced corrosion resistance [29,54,114–117], their effectiveness depends on the uniform dispersion of *h*-BN within the polymer matrix [32]. Achieving the maze effect and maintaining long-term barrier properties require uniform dispersion of 2D materials within the composite coating. However, the intrinsic hydrophobicity of *h*-BN, coupled with strong van der Waals interactions among *h*-BN flakes, often promotes aggregation, reducing their effectiveness as anticorrosion fillers when processed in water and polar solvents, as previously discussed in Section 2.1 [112].

The tendency of *h*-BN to aggregate in polar solvent-based paints due to its hydrophobic nature remains a critical challenge, requiring strategies to improve dispersion and uniformity within the polymer matrix. As explored in Section 2.3, the hydrophobicity of *h*-BN can lead to agglomeration of 2D flakes in polar solvent-based paints, impairing the physical barrier effect [108]. Additionally, the uniform dispersion of *h*-BN flakes is crucial for optimal anticorrosion performance of the coating [32,109]. Various methods have been employed to address these challenges and improve dispersion. Liquid-phase exfoliation (LPE), particularly wet-jet milling, has been effective in producing few-layer *h*-BN nanosheets with superior dispersibility. Surfactant-assisted dispersion, using stabilizers such as polyethyleneimine (PEI) [118–120] or polystyrene microspheres [63], improves interfacial compatibility and minimizes aggregation. Additionally, solvent selection plays a crucial role in maintaining dispersion stability, with polar solvents proving particularly effective [121,122]. Mechanical methods, such as ball milling, have also been explored [123,124], though excessive shear forces can introduce defects that compromise performance.

Functionalization of *h*-BN is a critical strategy to enhance its dispersibility, improve interfacial adhesion, and increase cross-linking density within polymer matrices, all of which contribute to superior anticorrosion coatings [125]. These strategies prevent agglomeration, strengthens polymer compatibility, and reinforces the coating's structural integrity, leading to improved long-term performance [83,84,126]. The functionalization methods can be broadly classified into non-covalent and covalent modifications. Non-covalent functionalization, often mediated by π-π interactions or surfactant-assisted dispersion, preserves the intrinsic properties of *h*-BN while improving its processability [127]. This approach includes polymer adsorption (*e.g.*, polyethyleneimine, polystyrene microspheres) and surface modifications using graphene oxide or carbon-based additives, which enhance compatibility with polymer matrices. On the other hand, covalent functionalization chemically modifies the *h*-BN surface, introducing functional groups that enhance both solubility and interfacial bonding [127]. Hydroxyl (-OH) functionalization [121,128,129] improves dispersion in polar solvents, while silanization (*e.g.*, TEOS, APTES) [58,128,130] increases compatibility with polymer coatings by strengthening interfacial adhesion. Additionally, polyaniline (PANI) grafting [131,132] has been employed to enhance the anticorrosion properties of *h*-BN composites. These functionalization techniques play a key role in optimizing *h*-BN-based coatings by preventing agglomeration, improving coating uniformity, and increasing hydrophobicity, thereby extending the durability and performance of protective layers [58,78,133–137].

Building upon these functionalization strategies, several studies have demonstrated their effectiveness in improving the dispersibility and anticorrosion performance of *h*-BN-based coatings. One approach involves modifying *h*-BN through organic functional groups to enhance interfacial adhesion and dispersibility. For instance, Cui *et al*. [64] improved the dispersibility of *h*-BN flakes by functionalizing them with carboxylated aniline trimer, resulting in enhanced anticorrosion performance. The latter was attributed improved water barrier properties of the epoxy coating, achieved by incorporating uniformly dispersed h-BN nanosheets [64]. A different strategy is to employ synergistic interactions with other 2D nanomaterials to enhance both dispersibility and protective performance. Wu *et al*. [84] non-covalently modified *h*-BN nanosheets with GO to improve the *h*-BN dispersion in a waterborne epoxy matrix. The strong π-π interactions between GO and *h*-BN led to improved dispersion, resulting in coatings with excellent barrier properties. The optimized 0.3 wt% GO/*h*-BN composite coating exhibited excellent corrosion resistance by reducing penetration pathways of corrosive agents [86]. Another promising approach involves the incorporation of boron nitride quantum dots (BNQDs) to further improve dispersibility and maximize the impermeability of *h*-BN-based coatings. Zhao *et al*. [138] functionalized *h*-BN nanosheets with BNQDs through strong π-π interactions and embedded them into a waterborne epoxy coating applied to Q235 steel-. Their results showed that even at low concentrations (0.1 – 0.5 wt%), functionalized BNQD coatings demonstrated significant improvements corrosion resistance due to better dispersibility and reinforcement of the polymer network [138]. Another strategy employed is polymer-stabilized functionalization. Wang et al. [126] employed polystyrene (PS) microspheres to non-covalently functionalize *h*-BN nanosheets through strong π-π interactions and produced *h*-BN-PS composite coating within a waterborne polyurethane (WPU) matrix through a facile latex blending method. This coating remarkably increased the impedance by 4 orders of magnitude compared to bare WPU coatings after a 28-day long immersion test, highlighting the enhanced dispersibility of the *h*-BN nanosheets in the polymeric matrix and consequently the corrosion resistance [126]. While non-covalent functionalization effectively enhances dispersion and prevents aggregation, its effectiveness in maintaining long-term interfacial bonding depends on the specific interactions employed. Approaches such as π-π interactions (*e.g.*, with graphene oxide or boron nitride quantum dots) can reinforce polymer compatibility, while polymer-assisted functionalization (*e.g.*, polyethyleneimine or polystyrene microspheres) introduces stability that prolongs dispersion uniformity [83,84,126].

Besides non-covalent modifications, covalent functionalization of h-BN nanosheets has been studied to enhance coating performance by improving interfacial bonding and chemical stability within polymer matrices [84,128,131,137,139]. The major challenge in this approach is ensuring functionalization does not disrupt the *h*-BN lattice, which is crucial for maintaining its barrier properties [54,56]One effective strategy is polymer grafting, where covalent interactions between *h*-BN and polymeric species enhance coating uniformity and resistance to corrosive agents. Tang *et al*. [131] fabricated *h*-BN/polyaniline (PANI) nanocomposites by *in situ* polymerizing aniline onto *h*-BN nanosheets. demonstrating that 2 wt% *h*-BN/PANI nanocomposites provided the best corrosion resistance. Their waterborne epoxy coatings exhibited no visible corrosion regions after 28 days of immersion in a 3.5 wt% NaCl solution, whereas coatings with only *h*-BN, PANI, or epoxy showed complete corrosion [131]. The synergistic effect of PANI not only improved *h*-BN dispersion but also extended the diffusion pathways of corrosive agents, thereby enhancing the protective performance of the coating [131]. Beyond polymer grafting, surface functionalization through oxidation and silane coupling has been developed to improve adhesion between *h*-BN and polymer matrices. Li *et al*. [128] oxidized *h*-BN with acid treatment and subsequently grafted silane onto the surface through a condensation reaction, which significantly improved the interfacial bond between the *h*-BN sheets and the epoxy resin. This modification effectively eliminated the micropores and cracks in the coating, reducing the corrosion current and increasing the polarization resistance by an order of magnitude compared to unmodified *h*-BN/epoxy coatings [128]. Another promising approach involves amine-based modifications to improve aqueous dispersibility and stability. Wu *et al*. [140] functionalized *h*-BN nanosheets using water-soluble branched polyethyleneimine through Lewis acid-base interactions, achieving homogeneous dispersion in a waterborne epoxy coating. This strategy significantly enhanced the anticorrosion performance on mild steel during immersion tests, demonstrating the potential of covalent amine functionalization in improving both dispersion and adhesion in polymer matrices. Additionally, Wan *et al*. [139] explored carbon dot functionalization as an alternative means to improve dispersibility and interfacial bonding of the *h*-BN nanosheets within WEP matrix. The introduction of carbon dots improved the stability of *h*-BN within the coating and resulted in long-term protection, maintaining corrosion resistance for 40 days in saltwater immersion tests [139]. These studies collectively demonstrate that covalent functionalization not only mitigates *h*-BN aggregation but also enhances its adhesion, chemical stability, and overall coating performance. In applications where strong interfacial bonding and long-term chemical stability is required, covalent strategies often provide enhanced corrosion protection compared to non-covalent modifications [128,131,139,140].

In addition to improving dispersibility, functionalization also plays a crucial role in reinforcing polymer cross-linking density. By modifying interfacial adhesion and promoting stronger polymer bonding, functionalized *h*-BN ensures a uniform distribution within coatings while simultaneously reducing the presence of microvoids and cracks that serve as diffusion pathways for corrosive agents [83,125,140,141]. Increased cross-linking not only enhances the durability of the coatings but also reduces the permeability of corrosive species, thereby extending their service life [83,125,140,141]. Several studies have demonstrated the effectiveness of different functionalization strategies in enhancing cross-linking and adhesion. Kulia *et al*. [137] functionalized exfoliated *h*-BN nanosheets using tetraethyl orthosilicate (TEOS) and developed *h*-BN-reinforced epoxy composite coatings with different filler loadings (0-15 wt%). Their results showed that at an optimal 5 wt% filler loading, the coating exhibited significantly improved anticorrosion performance due to enhanced cross-linking in the epoxy matrix, which reinforced the polymer structure and minimized permeability [137]. Similarly, silane-functionalized *h*-BN nanosheets in epoxy coatings were shown to increase both thermal conductivity and anticorrosion performance at filler contents up to 15 wt% with improved cross-linking density compared to coatings with lower filler contents [58]. In addition to silane coupling, carbon dot functionalization has emerged as a promising route to reinforce cross-linking. Wan *et al*. [139] demonstrated that carbon dot decoration of *h*-BN nanosheets reduces aggregation by intercalating between the nanosheets. improving dispersibility and their compatibility with the polymer. The enhanced dispersion facilitated stronger polymer interactions, leading to more uniform cross-linking, which in turn strengthened the structural integrity of the coating and prolonged the diffusion pathways of corrosive agents, significantly improving overall barrier performance [139].

Beyond improving cross-linking, certain functionalization approaches provide additional corrosion inhibition mechanisms. Yuan *et al*. [142] decorated *h*-BN surfaces with ZnO and pDA to create a coating for Al surfaces. This dual-functionalization not only improved the dispersibility and uniformity of *h*-BN flakes within the epoxy matrix but also provided active corrosion inhibition ability through anion capture mechanisms [142]. The release of pDA and $Zn^{+2}$ ions released in response to pH change provided additional protection [142]. Polydopamine molecules migrated to the anode region to anchor $Al^{+3}$ ions, while $Zn^{+2}$ ions capture the $OH^-$ and $Cl^-$ ions, effectively suppressing the cathodic reactions and mitigating localized corrosion effects [142]. Moreover, released $Zn^{+2}$ ions captured anions, further enhancing the anticorrosion performance of the coating [142].

One approach to intrinsic self-healing relies on incorporating dynamic covalent bonds, which enable crack repair under external stimuli such as heat. Du *et al*. [54] developed ultrathin functional *h*-BN films via ball milling in an ionic liquid, achieving excellent dispersion in waterborne epoxy coatings. These well-dispersed *h*-BN nanosheets contributed to the "maze effect," reducing ion diffusion pathways, while the ionic liquid imparted self-healing properties. After a 4-week immersion test in 3.5 wt% NaCl, their functionalized *h*-BN/epoxy composite exhibited a coating resistance that was 3–4 orders of magnitude higher than pure epoxy [54]. Similarly, Cao *et al*. [143] leveraged Diels-Alder reactions to create a thermally reversible crosslinked epoxy coating using maleimide-functionalized *h*-BN nanosheets. When heated to 120°C, the coating demonstrated effective crack repair, regaining its anticorrosion efficiency with a corrosion inhibition rate of 97.2% at a 2% crosslinking degree. These studies highlight how *h*-BN-based coatings can be engineered for thermal-induced self-healing applications. Extrinsic self-healing strategies, on the other hand, rely on functionalized *h*-BN serving as a carrier for corrosion inhibitors, enabling controlled release when triggered by environmental changes such as pH shifts. Zou *et al*. [144] incorporated polydopamine (pDA)-coated h-BN nanosheets along with benzotriazole (BTA)-loaded β-cyclodextrin nanocapsules into an epoxy matrix. The pDA functionalization improved the dispersibility and barrier performance of *h*-BN, while BTA release was pH-responsive, allowing corrosion inhibition upon environmental acidification [144]. Their hybrid coating provided sustained corrosion protection over a 40-day immersion test, demonstrating the effectiveness of *h*-BN-based inhibitor delivery systems [144].

The hydrophobicity of protective coatings plays a crucial role in corrosion resistance by reducing water absorption and limiting electrolyte penetration. While *h*-BN itself is inherently hydrophobic, its impact on coating performance is influenced by its concentration, dispersion, orientation, and interactions with the polymer matrix. The exact threshold varies based on the polymer matrix properties, dispersion methods, and production techniques. Molina-Garcia et al. [29] found that *h*-BN produced via wet-jet milling reached maximum hydrophobicity at 10 wt% in a PIB matrix, but higher loadings led to aggregation. In contrast, other studies have reported agglomeration occurring at much lower concentrations, typically between 0.5–2 wt%, depending on the matrix and processing conditions [55,83]. Beyond concentration, the alignment and distribution of *h*-BN nanosheets within a polymer coating can significantly affect the overall water resistance, as randomly oriented flakes create micro-roughness and air pockets, promoting superhydrophobic behaviour, whereas poor dispersion can lead to defects that allow water infiltration [78,84,114]. Several studies have explored the strategies to enhance the hydrophobic performance of *h*-BN-based coatings through structural modifications and hybrid compositions. Li *et al*. [129] incorporated polytetrafluoroethylene (PTFE) and hydroxylated *h*-BN (OH-*h*-BN) into an epoxy matrix, achieving a water contact angle of 120° due to the combined effects of low surface energy of PTFE and barrier properties of *h*-BN. This composite exhibited five orders of magnitude higher low-frequency impedance after 80 days of immersion in 3.5 wt% NaCl solution, demonstrating long-term durability [129]. Similarly, Zhang *et al*. [117] modified *h*-BN nanosheets with $Fe_3O_4$ nanoparticles, which promoted an ordered nanosheet structure within the polymer, increasing the tortuosity of diffusion pathways for corrosive agents. This structure enhanced water resistance by limiting moisture penetration through the coating [117]. Sun et al. [78] further demonstrated that mussel-inspired urushiol-functionalized BNNSs exhibited improved dispersion and enhanced hydrophobicity, significantly increasing the impedance of coatings after 50 days of saltwater immersion. Functionalized *h*-BN nanosheets are also used to make superhydrophobic coatings, as well. Wang *et al*. [135] modified *h*-BN nanosheets with benzoxazine, achieving a superhydrophobic coating with a contact angle of 158.2°. Even after harsh environmental exposure and a 30-day immersion test in 3.5 wt% NaCl solution, the coating maintained its superhydrophobic properties, demonstrating excellent long-term water repellency [135]. Wu *et al*. [84] used GO to functionalize the *h*-BN nanosheets, which were then incorporated within waterborne epoxy.

While GO improved the compatibility and dispersibility of *h*-BN nanosheets with epoxy, *h*-BN provided hydrophobicity to the coating beyond physically blocking the corrosive agents [84].

The synergy between *h*-BN nanosheets and polymers not only enhances anticorrosion properties but also improves other key qualities of coatings, such as wear resistance, mechanical strength, and thermal stability, due to the high mechanical strength and chemical stability of *h*-BN. Kumar *et al.* [145] demonstrated that $TiO_2$-decorated *h*-BN nanosheets in PEDOT coatings applied to 316L stainless steel implants improved resistance to pitting corrosion and wettability, indicating promising potential for biomedical applications. Similarly, Yan *et al.* [133] investigated oxidized carbon fiber (OCF)-functionalized *h*-BN-reinforced epoxy coatings for heat exchangers. Their study revealed that oxidation treatments on OCF significantly enhanced thermal conductivity and corrosion resistance, though the introduction of hydrophilic carboxyl groups negatively impacted coating performance by increasing water uptake. This emphasizes the necessity of balancing hydrophilicity and hydrophobicity through appropriate functionalization strategies to optimize performance [133]. Beyond improving corrosion resistance, *h*-BN-based coatings can also mitigate wear and crack propagation in protective layers. The functionalization of *h*-BN nanoplatelets through π-π interactions of amine-capped not only improved anticorrosion properties but also reduced porosity and crack propagation while increasing wear resistance and lowering the friction coefficient of the coating [83]. Zhao *et al.* [146] further demonstrated that phosphorylated *h*-BN nanosheets, synthesized by exfoliating hydroxylated *h*-BN followed by phosphate esterification with phosphoric acid, created a highly effective barrier against corrosive ions. When applied to carbon steel, the phosphorylated *h*-BN significantly inhibited ion diffusion, reducing the overall corrosion rate in 3.5 wt% NaCl solution [146].

## 4. Industrial Exploitation

Industrially viable pathways for corrosion protection of metallic structures and components include cathodic protection, surface treatments, coatings and materials engineering through alloying. Among these, as discussed in the previous sections, coatings, and in particular organic coatings, can strongly benefit from the peculiar properties of *h*-BN nanosheets [20,22,47,147–149]. Normally, a coating cycle designed to provide protection in marine environments is either made of high-thickness epoxy coatings or a combination of different layers with specific properties [150]. These layers mostly include a zinc-rich primer or a metallic layer (Zn or Al), an epoxy or acrylic base coat aimed at providing coverage and thickness and a polyurethane topcoat for wear and UV resistance and aesthetics. In the case of thick epoxy coatings, the protection mechanism only involves a strong barrier effect given by the compactness of the polymeric matrix and the long diffusion path required for corrosive species to reach the substrate. Furthermore, the deposition of large amounts of material also helps the operators in avoiding defects during the application.

In this context, the incorporation of *h*-BN nanosheets can significantly improve the coatings resistance to diffusion. This is achieved by increasing the tortuosity of the diffusion pathways within the coating and enhancing the overall compactness of the composite compared to the unmodified polymer. This would enhance the corrosion protection performance of the coating or enable the use of a thinner layer, thereby reducing material consumption, VOC emissions, and deposition time [151]. Additionally, thinner coating layers typically offer better mechanical properties and are less prone to cracking.

Considering the alternative approach of applying a multi-layer coating system, including Zinc -rich primers, its main drawbacks include prolonged deposition and drying times, as well as the environmental impact from the presence of Zn particles [152], which could dissolve into seawater if the coating is damaged. Also, the adherence of the topcoat to Zinc-rich primer must be carefully addressed to avoid insufficient adhesion strength causing coating delamination.

The use of *h*-BN nanosheets as functional fillers in any of the two non-conductive layers, *i.e.,* base coat and topcoat, can improve their barrier properties and eventually achieve outstanding performances also without a cathodically active primer, provided that the coating system is formulated to have a good adhesion to the substrate [29]. This would solve the problem of the environmental impact of Zn particles [152], without compromising the other properties of the coating. Furthermore, *h*-BN has good hiding power (*i.e.*, ability of a paint or pigment to hide the surface that the paint was applied to), and can be used in aesthetically relevant layers without significant modifications. Finally, *h*-BN-loaded composites have good mechanical performances and wear resistance, making *h*-BN nanosheets a perfect choice for this application.

In the industrial world, regulations and standards determine guidelines and obligations for the choices of materials and procedures. Furthermore, these documents give an insight into the performance range allowed for the installation of a product. For example, coatings for offshore wind turbines, which are subjected to different corrosive environmental conditions (ranked from C1 to C5 and CX in the case of atmospheric corrosion, according to ISO 12944-2), must comply with many standards to be allowed for installation. Although being present on the whole structure, due to the extreme conditions offshore installations are subjected to, other corrosion protection mechanisms (mostly cathodic protection) must be used to achieve an acceptable durability of the structure and good safety conditions for workers. ISO standards regulate both the type of coatings to be applied to the different parts subjected to different environmental conditions and their performance.

The same ISO 12944 defines the allowed coating cycles, specifying the type of paint and the thickness for each layer, together with the required adhesion strength and the foreseen thickness loss and durability range. As mentioned in the previous paragraph, these coating systems usually comprise a cathodically protective primer (*i.e.*, Zn-rich organic or inorganic coating, thermally sprayed Zn or Al, hot dip galvanizing) deposited on a surface prepared according to precise specifications. After the primer, several layers of organic barrier paint must be applied with a dry final thickness ranging from 250 μm to 800 μm [29]. Other layers can then be applied according to the specific environmental conditions. Of course, other applications in different environments would require different specifications for the coating system, usually less demanding.

Standards also help in the definition of tests for the evaluation of the performance of coatings. In a specific study followed by some authors in collaboration with an industrial partner in the framework of the Municipality of Genoa project (*i.e.*, PFBP), industrially produced few-layer *h*-BN flakes (BeDimensional S.p.A.) were used as an anticorrosion filler to formulate a novel coating system for the protection of structural steel parts. Besides resistance to a highly corrosive environment, the structural steel structure also needs mechanical and abrasion protection. For this last reason, common protective coating cycles could not be applied, and a more mechanically strong layer was designed. The final coating system comprised an inorganic Zn-rich primer and a proprietary acrylic topcoat loaded with *h*-BN flakes [29]. The effect of loading on the anticorrosive performance of the coating was studied through electrochemical measurements according to the ASTM G5 standard on steel samples coated with the sole acrylic topcoat [29]. Tafel plots and corrosion rates are reported in Fig. 4 and show a characteristic behavior which is reported in Ref. [29], *i.e.*, an optimum value of concentration of *h*-BN, which guarantees the best anticorrosion performance.

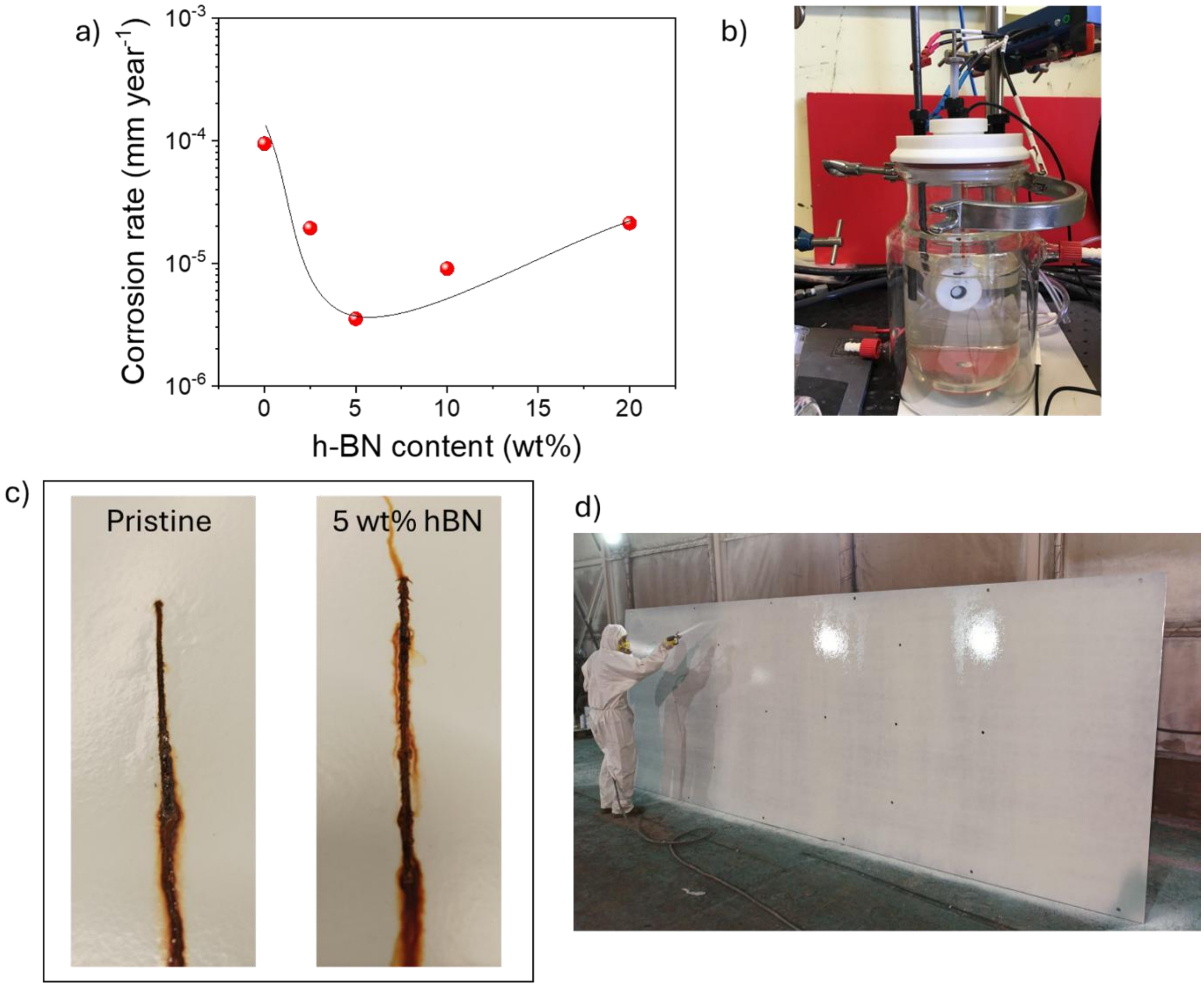

***Figure 4.*** *a) Corrosion rate (according to ASTM G5) of structural steel coated with a proprietary acrylic topcoat loaded with different amounts of h-BN nanosheets. b) Electrochemical setup for corrosion measurements according to ASTM G5. c) Steel substrates coated with pristine acrylic topcoat and topcoat loaded with 5 wt% h-BN nanosheets after 1000 h salt spray fog test according to ISO 9227. Both samples were previously coated with a Zn-rich inorganic primer. d) Photo of the deposition of the h-BN-based proprietary acrylic topcoat on a large area 12m$^2$ the structural steel structure.*

Furthermore, other industrially meaningful characterization techniques can be used to study the performance of the complete coating system in a real-world application. In the context of the aforementioned project, the resistance to corrosion was evaluated through neutral salt spray test (ISO 9227) and immersion test in sea water (ISO 2812-2), reaching a resistance of more than 1000 h for the first and more than 24 months for the latter. According to the reported observations, the addition of *h*-BN nanoflakes to the topcoat improved corrosion resistance by reducing the amount of blistering and delamination around the scribe, thus providing higher durability in case of mechanical damage to the coating. In industrial applications it is mandatory that the proposed anti-corrosion coating systems fully satisfy the requirement and indications given by international standards before the coating can be validated commercially. Some of the main tests to fully characterize the coating from the electrochemical and mechanical point of view are reported in **Table 1.**

*Table 1: List of international standards for the mechanical and electrochemical validation of anticorrosion coatings.*

| Test | Standard |
|---|---|
| Salt-spray fog accelerated corrosion | ISO 9227 |
| Sea water immersion | ISO 2812-2 |
| Pull-off adhesion | ISO 4624 |
| Bending | ISO 1519 |
| Surface hardness – Shore<br>Surface hardness – Buchholz | ISO 7619-1<br>ISO 2815 |
| Impact resistance | ISO 6272 |
| Abrasion resistance | ISO 7784 |
| Sand abrasion resistance | ASTM D968 |

As a result of these efforts, several patents in this field have already been filed/published globally, including those from some of the authors of this review (IT102021000013169; PCT/IB2022/054717, IT102023000018468 and 102024000001095), testifying the upgrowing interest in technology. Lastly, one main concern regarding industrial applications that must be carefully addressed is that of material and manufacturing costs. Provided that, according to literature and patents, the processing of the paint formulation can be easily upscaled without substantial modifications to the standard processes, material cost remains a relevant issue in the case of nanomaterials, such as *h*-BN nanoflakes. Taking bottom-up synthesis approaches out of the equation due to their low productivity and high energy costs (controlled environments and high temperatures are mostly needed), some top-down techniques appear to be scalable for an affordable mass production [42,43]. Among those, LPE [153,154] is, to the knowledge of the authors, the most suitable method for massive production of high-quality *h*-BN nanosheets (see, for example, WO2017060497A1 and WO2017089987A1), mostly thanks to its versatility and ability to process all kinds of layered materials [29,42,43,112]. Further development of the production methods is currently promoting the use of *h*-BN in widespread applications.

## 5. Outlook

The exploration of *h*-BN nanosheets as a functional filler for anticorrosion coatings offers significant promise for advancing in materials science, particularly for enhancing structural integrity and providing long-term corrosion resistance. *h*-BN stands out due to its unique properties, including its exceptional chemical stability, impermeability, and electrical insulation, distinguishing it from other established 2D materials such as graphene. These advantages make it well-suited for industrial applications in which mitigating galvanic coupling and managing electrochemical interactions are critical issues.

However, several key challenges remain before *h*-BN nanosheets can be widely integrated into commercial applications. The production of high-quality, defect-free *h*-BN nanosheets at an industrial scale remains a significant hurdle. Current fabrication methods often introduce structural imperfections, such as wrinkles and grain boundaries, which compromise its barrier properties. Overcoming these issues will require advances in both synthesis techniques and the functionalization of *h*-BN nanosheets to enhance their dispersibility and compatibility with polymer matrices.

Looking ahead, future research should focus on optimizing *h*-BN-based composites, refining functionalization and dispersion techniques to minimize agglomeration. The integration of self-healing properties into *h*-BN-based coatings represents another exciting frontier, potentially extending the lifespan of these coatings, especially in harsh environments. Additionally, investigating new hybrid coating systems that combine *h*-BN with other 2D materials or polymers may offer new opportunities to enhance both corrosion resistance and mechanical properties.

The potential for *h*-BN nanosheets to replace or complement existing coating technologies is particularly compelling in industries in which environmental and safety regulations increasingly restrict the use of heavy metals such as zinc. As scalable production methods for *h*-BN continue to evolve, *h*-BN-based coatings could play a crucial role in the development of sustainable, high-performance anticorrosion solutions across sectors such as marine, energy, and beyond.

**Acknowledgement**

This project has been supported is conducted under the REDI Program, a project that has received funding from the European Union's Horizon 2020 research and innovation programme under the Marie Skłodowska-Curie Grant Agreement No. 101034328. This paper reflects only the author's view, and the Research Executive Agency is not responsible for any use that may be made of the information it contains. ICN2 acknowledges the Grant PCI2021-122092-2A funded by MCIN/AEI/10.13039/501100011033 and by the 'European Union NextGenerationEU/PRTR'. BeDimensional acknowledges funding from the European Union's DIAMOND Horizon Europe research and innovation program under Grant Agreement No. 101084124, the European Union's LAPERITIVO Horizon Europe research and innovation program under Grant Agreement No. 101147311, from the Italian Ministry of Environment and Energy Security in the framework of the Project GoPV (CSEAA_00011) for Research on the Electric System, and from the Italian Ministero della Università e Ricerca (MUR) in the framework of the Project FIT4H2. BeDimensional acknowledges PFBP project partners.